# Emergence of directionally-anisotropic mobility in a faceted Σ11 <110> tilt grain boundary in Cu

Megan J. McCarthy [a], Timothy J. Rupert [a, b, *]

[a] Materials and Manufacturing Technology, University of California, Irvine, CA 92697, USA

[b] Department of Materials Science and Engineering, University of California, Irvine, CA 92697, USA

* Email: trupert@uci.edu

## ABSTRACT

Faceted grain boundaries, where grain boundary area is increased in the name of producing low-energy segments, can exhibit new and unexpected migration trends.  For example, several faceted Σ3 boundaries have demonstrated anti-thermal and thermally damped mobility.  Σ11 <110> tilt boundaries represent another promising but relatively unexplored set of interfaces, with a (113) low-energy plane that can lead to faceting.  In this study, molecular dynamics simulations are used to study grain boundary migration of an asymmetric Σ11 <110> grain boundary in two face centered cubic metals.  Mobility of this boundary in Cu is strongly dependent on the direction of the applied driving force.  The mobility anisotropy generally becomes smaller, but does not disappear completely, as temperature is increased.  In contrast, the same boundary in Al demonstrates similar mobilities in either direction, illustrating that the anisotropic mobility phenomenon is material-dependent.  Finally, relationships between stacking fault energy, facet junction defect content, and boundary crystallography are uncovered that may inform future studies of faceted grain boundaries.

# 1. Introduction

Recent research has revealed that faceted grain boundaries can have unusual properties that may dramatically affect grain growth [1]. Faceted grain boundaries are interfaces that dissociate from one flat interface into two planes with different energies, one high and one low. The net energy of the boundary is lower in the faceted configuration than in the original flat form, despite the existence of additional defects (facet junctions) and increased grain boundary area. While these structures are themselves highly interesting and have been a topic of research for many decades [2-6], the effect of faceting on boundary properties such as mobility has only recently received attention and it is likely that faceted boundaries can dramatically influence microstructural evolution. In situ transmission electron microscopy heating experiments by Merkle et al. [7] revealed that faceted sections of a low-angle Au boundary were significantly less mobile than a similar faceted boundary in Al. Holm and Foiles [8] showed that even a small fraction of smooth boundaries (boundaries that do not undergo a roughening transition at high temperatures) can stagnate grain growth in pure Ni. Grain growth could be similarly slowed by introducing a population of low-mobility faceted boundaries into a microstructure. Faceted boundaries have also been implicated in abnormal grain growth by Lee et al. [9], who observed a defaceting transition that was triggered above a homologous temperature ($T_H$) of 0.7. An understanding of the mechanisms behind these types of unique boundary mobilities may shed new light on grain growth phenomena, making the study of faceted boundary migration warranted.

Faceting transitions and grain boundary migration are inherently atomistic processes, making atomic scale modeling extremely useful for studying such behavior. For example, many studies have used molecular dynamics to probe the mobility of symmetric and asymmetric tilt and twist boundaries [10-14]. However, only a few literature reports have focused on the mobility of faceted boundaries [1, 15-17]. The studies that do exist for faceted boundaries has already shown them to exhibit unusual mobility trends. For example, Humberson and Holm [1] reported anti-thermal grain boundary mobility for a $\Sigma3$ <111> 60° {11 8 5} boundary in Ni. While grain boundary migration is typically a thermally-activated process where higher temperatures lead to higher mobilities, these authors actually found that the interface's mobility decreased as temperature was increased, with migration occurring via the coordinated motion of Shockley partial triplets located in the high-energy facet plane. Experiments by Priedeman et al. [16] showed that nano-faceted $\Sigma3$ boundaries along the <110> tilt axis exhibit thermally damped behavior. One reason for this relative lack of studies on faceted boundary mobility is their structural complexity. Each sub-structure in a faceted boundary plays its own role in boundary migration, and those structures can interact with each other in complex ways [15]. This leads to the practical limitation that larger computational models must typically be used to probe a faceted boundary compared to that needed to study standard, planar interfaces.

The understanding of boundary faceting can be broadened by looking beyond $\Sigma3$ grain boundaries to investigate other low-index coincident site lattice (CSL) interfaces, with the $\Sigma11$ <110> tilt grain boundary in particular standing out as a potentially interesting system. The static structures of various symmetric and asymmetric $\Sigma11$ boundaries have been studied in prior work [2, 3, 5, 18-22], providing a rich knowledge base showing that faceting is common. For example, Brown and Mishin [3] showed that the high energy segments of faceted $\Sigma11$ boundaries are often oriented along an unusual plane, the {001}/{111} interface, which is not a member of the $\Sigma11$ CSL boundary set. We hypothesize that the asymmetric geometry of these boundaries may lead to unusual mobility trends, and there appear to be no published studies of them to our knowledge. In addition, Panzarino and Rupert [23] measured an increase in $\Sigma11$ boundary fraction, including a number of faceted Σ11 boundaries, after a molecular dynamics study of cyclic deformation in nanocrystalline Al. This finding demonstrates that faceted Σ11 boundaries can also be created by mechanically-driven grain boundary migration in nanocrystalline materials.

In this study, we report on the grain boundary migration and mobility of faceted Σ11 boundaries using an artificial driving force (ADF) method. A Σ11 <110> bicrystal is probed in two face centered cubic materials (Cu and Al) at three different temperatures. We find that grain boundary migration is highly directionally-anisotropic in Cu, meaning that mobility is not the same for a boundary moving in opposite directions. The anisotropic mobility appears to be affected by the local boundary structure and the simulation temperature. After detailed analysis of boundary migration for this system, we conclude that a combination of boundary asymmetry and low stacking fault energy gives rise to a directionally-favored motion mechanism (slip plane shuffling), which in turn leads to directionally-anisotropic grain boundary mobility.

## 2. Methods

The geometry of a bicrystal can be defined by five angles, which represent the macroscopic degrees of freedom. The first three, called the misorientation, represent the rotations needed to bring the two crystals into coincidence and determines the CSL value of the bicrystal (e.g., Σ11). The other two, called the boundary plane orientation, determine the direction of the boundary plane's normal vector. The angles of the boundary plane orientation include the azimuthal angle, α, and the inclination angle, β (sometimes called the polar angle). In tilt boundaries, α is parallel to the tilt axis and is thus by definition 90°. The inclination angle β has a range of $0° \leq \beta < 90°$ (a result of four-fold symmetry in cubic structures).

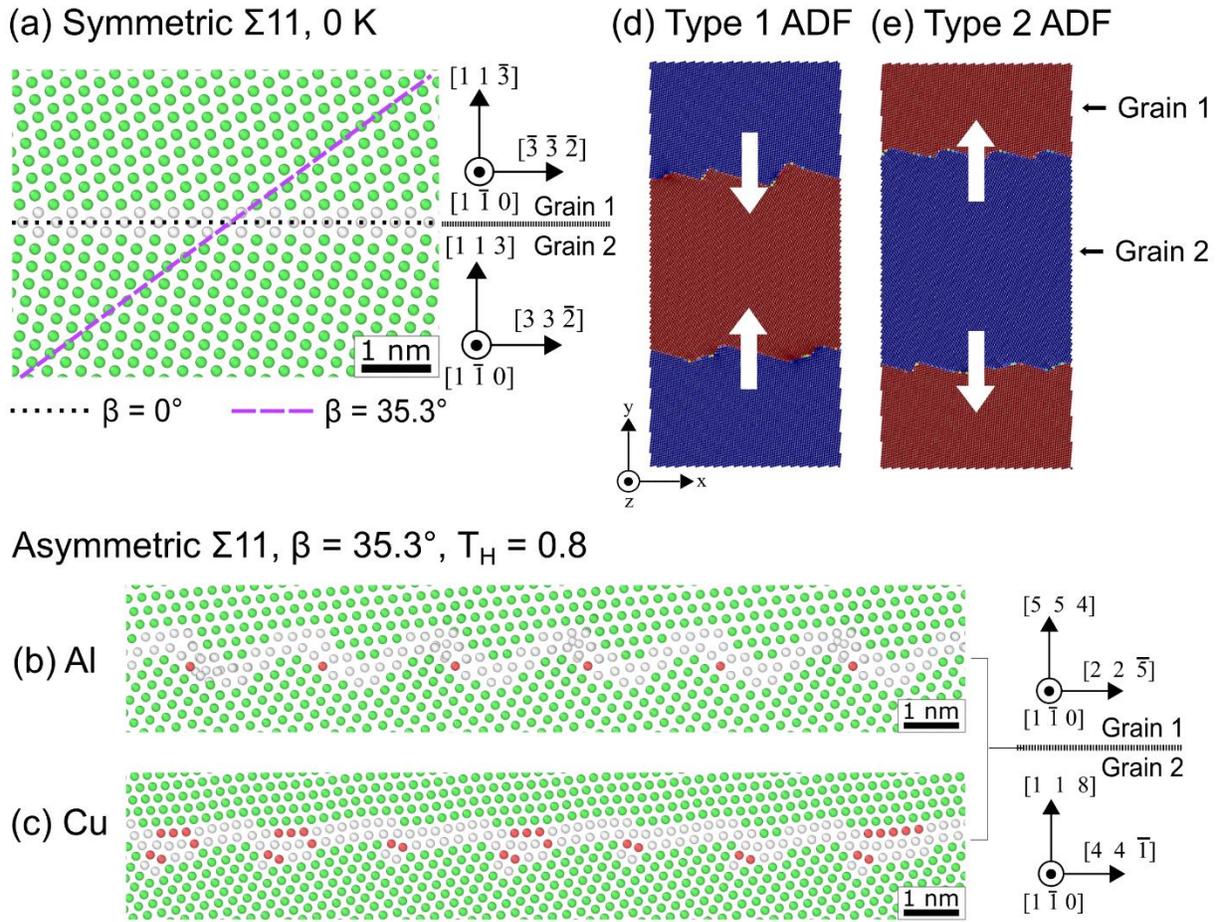

Figure 1. (a) The Σ11 symmetric β = 0° boundary, indicated by a black dotted line, which has the lowest energy of any Σ11 boundary. The boundary plane with β = 35.3° under study here is indicated by the purple dashed line. (b, c) The as-annealed boundaries at $T_H$ = 0.8 for β = 35.3° in (b) Al and (c) Cu, which have identical crystallography (shown on the right side of both boundaries), where different potentials have different facet morphologies. (d, e) The Type 1 and Type 2 ADF, where the favored (blue) and unfavored (red) grains are swapped to change the direction of boundary migration.

By varying β, one may produce a wide array of tilt bicrystal structures. The one with the lowest-energy and highest symmetry is the symmetric boundary plane (SBP) that is found at β = 0°. The SBP for the Σ11 <110> tilt bicrystal is shown in figure 1(a) (note that the viewing angle of all boundary images in this work is down the <110> axis), which has an orientation of (113)₁/(11-3)₂. Faceting is predicted for this type of boundary when β is less than approximately 60° [3]. For this study, we chose β = 35.3°, corresponding to the purple dotted line in figure 1(a).

| EAM | $a_0$ [Å] | $T_{melt}$ [K] | $\gamma_{SF}$ [mJ m$^{-2}$] | E Σ11, 0° [mJ m$^{-2}$] | E Σ11, 35.3° [mJ m$^{-2}$] | E IBP* [mJ m$^{-2}$] | E Σ5, 32.5° [mJ m$^{-2}$] |
|---|---|---|---|---|---|---|---|
| Al [24] | 4.05 | 1035 | 146 [24] | 150.8 | 352.2 | 270.9 | 526.8 |
| Cu [25] | 3.615 | 1357 | 44 [25] | 309.9 | 608.7 | 436.8 | 986.3 |
| * IBP energy estimated from 80% of the 29.5° boundary energy [3] | | | | | | | |

Table 1. Selected properties of the potentials, Σ11 boundaries, IBP, and Σ5 boundaries, including grain boundary energies (E) and stacking fault energy ($\gamma_{SF}$).

An asymmetric Σ5 <001> tilt boundary (β = 32.5°) was also generated and probed to provide a boundary that is asymmetric without faceting for comparison. Fully periodic simulation cells were generated in the Large-scale Atomic/Molecular Massively Parallel Simulator (LAMMPS) [26], using code developed by Tschopp et al. [5] for the identification of minimum energy grain boundary structures. This algorithm uses a series of iterating shifts and atom deletions to probe all possible periodic structures for the given crystal orientations and also calculates grain boundary energy. From the set of generated cells, the lowest energy option is selected.

Visualization of boundaries and parts of the data analysis were performed using the OVITO software toolset [27]. Boundary snapshots are quenched using a conjugate gradient technique that removes thermal noise to allow for detailed structural analysis and then colored according to common neighbor analysis [28], with green indicating local face centered cubic (FCC) orientation, red indicating hexagonal close packed (HCP) orientation, and grey indicating any other or an undetermined orientation. Two EAM potentials were utilized, with one representing Al [24] and the other modeling Cu [25]. Various properties of the potentials, as well as grain boundary energies calculated from the Tschopp algorithm described above, are included in Table 1. The melting temperatures for each potential were confirmed to be within ±5K of the reported values using the method outlined by Wang et al. [29].

At different points in this work, we refer to these two EAM potentials as different "materials." When using this term, it is crucial to acknowledge that interatomic potentials are attempting to simulate atomic interactions and are not always capable of replicating all of the properties of the named element or alloy. Especially with dynamic behavior such as grain boundary mobility, there can be great variability between potentials representing the same element or material [17]. However, the phenomenon of directionally-anisotropic mobility is shown to be strongly related to stacking fault energy, which correlates with grain boundary dislocation content and faceted $\Sigma 11$ structure. For the Cu potential, the calculated stacking fault energy of 44 mJ m$^{-2}$ shown in Table 1 agrees very well with the reported experimental value of 45 mJ m$^{-2}$ [25]. The Al potential's stacking fault energy of 146 mJ m$^{-2}$ [24] is an intermediate value within the experimentally-measured range of 120-166 mJ m$^{-2}$. Therefore, both potentials describe the stacking fault energies well and also represent the differences between the two materials that are important with respect to structure (and thus also directionally-anisotropic mobility). For these reasons, we use the term "material" to describe simulations that use the different interatomic potentials that are approximating atomic interactions in Al and Cu.

System size is an important consideration for grain boundary mobility studies [1, 12, 14, 30, 31]. Bicrystals that are too small, specifically those that are too short in the direction of the grain boundary normal, have resulted in problematic mobility artifacts [12, 14, 31]. Other recent work has shown that these size considerations may only strictly apply to systems with flat boundaries, and less so to asymmetric or defect-heavy boundaries such as the asymmetric $\Sigma 11$ and $\Sigma 5$ boundaries studied here [32]. However, in this study, we conservatively choose to ensure that the boundary is large enough to avoid any potential issues. For both the $\Sigma 11$ and $\Sigma 5$ boundaries, the total height perpendicular to the grain boundary normal, $L_y$, was made equal or greater than 30 nm to alleviate any concerns. The generated $\Sigma11$ bicrystals had heights of 33.4 nm for Al and 35.8 nm for Cu, within the acceptable zone outlined by Deng and Deng [14]. The minimum tilt-axis thickness, $L_z$, was fixed to be ten repeats of the lattice parameter, $a_0$, giving thickness values of 4.1 nm and 3.6 for Al and Cu, respectively. The minimum length for $L_x$, the direction parallel to the grain boundary plane, was set to 9 periodic repeats of the generated facet structure, producing a length of 21.2 nm in Al and 19.1 nm in Cu. The total number of atoms in each simulation cell was 166,068 for Al and 198,576 for Cu. Both the $\Sigma5$ in Al and in Cu had 143,400 atoms, with Al measuring 17.4 nm by 34.8 nm by 41.0 nm ($L_x$, $L_y$, $L_z$), and Cu measuring 15.6 nm by 31.2 nm by 36.9 nm ($L_x$, $L_y$, $L_z$).

Once generated, boundaries were relaxed using an NPT ensemble with a Nosé-Hoover thermostat and barostat to regulate temperature and ensure zero pressure on the cell. Annealing runs were initialized by adding a randomized velocity to atoms in the minimized bicrystals corresponding to half the target temperature. Heating was then applied for 120 ps, with a ramp to bring the temperature to its target value within approximately the first 10 ps. Three simulation temperatures were chosen, corresponding to homologous temperatures of approximately 0.8, 0.85, and 0.9. Examples of as-annealed $\Sigma11$ boundaries for each potential at $T_H = 0.8$ are shown in figure 1(b) and (c), demonstrating that faceting occurs for both Al and Cu but the local structure is different. These structures will be discussed more thoroughly in the Results and Discussion section. At least 6 equivalent configurations were created for each combination of temperature, potential, migration direction, and CSL, using unique, randomly-generated velocity seeds for the initial temperature.

After equilibration at the target temperature, the energy-conserving orientational artificial driving force (ADF) developed by Ulomek et al. [33] was applied for a minimum of 120 ps and up to 1 ns for very slow/immobile boundaries. A cutoff value of 1.1 $a_0$ was chosen to capture first and second nearest neighbors. The ADF functions by adding an orientation-dependent energy to one grain of the bicrystal (i.e., the 'unfavored' grain). This added energy creates an energy gradient across the interface with the second, 'favored' grain. To lower the system energy, the atoms in the unfavored grain move to orient themselves with the favored grain's lattice, leading to grain boundary migration. In this study, the choice of the favored and unfavored grain is swapped to observe differences in boundary migration direction. Thus, a clear designation of which grain is the favored or growing grain is important. For the remainder of this work, we refer to the growth of Grain 1 as 'Type 1' behavior and growth of Grain 2 as 'Type 2' where figures 1(d) and (e) show the two different options. The blue-colored region is the favored grain in these figures and it is growing in the direction of the white arrows.

Driving force values in the range of 10-25 meV/atom were initially tested. Because some boundaries at the lowest homologous temperature of 0.8 are relatively immobile, the higher value of 25 meV/atom was used. This choice is reasonable because the boundaries studied here are non-planar and have a high defect content. Several prior studies indicate that both of these features can strongly impact a boundary's sensitivity to high driving forces [12, 31, 32]. In addition, Race et al. [32] found that a driving force of 25 meV/atom, the same as was used here, did not alter the fundamental migration mechanisms of heavily defected boundaries, including a faceted boundary. In addition, molecular dynamics studies of $\Sigma 3$ <110> tilt boundaries have shown that especially slow moving boundaries require higher driving forces for appreciable motion [30, 34]. Grain boundary velocities, $v$, were measured by tracking the mean position of each of the two boundaries separately for at least 50 ps of steady-state motion. Mobility, $M$, was then calculated as:

$$M = \frac{v}{P},\qquad(1)$$

where $P$ is the pressure experienced by the boundary (in this case, through the artificial driving force). The units of $M$ are m GPa$^{-1}$ s$^{-1}$.

## 3. Results and Discussion

### 3.1 Detailed Boundary Structure

We begin with a detailed description of the equilibrium structures of the boundaries in Al and Cu in figures 2 and 3, respectively. In addition to atomic snapshots of each boundary, these figures contain schematics highlighting local structural units, relevant crystallographic planes, and other important features. The shading of each atom indicates the approximate plane height with respect to the tilt axis, with darker atoms being one {110} plane height lower than lighter ones. In order to simplify the characterization of boundary structure, we will utilize a common tool for identifying grain boundary structures, the structural unit model (SUM) [35].

Two facets of the as-annealed and quenched Σ11 boundary in Al are shown in figure 2(a), with brackets indicating the location of each. They have a clear faceted shape, with significant variations in the location of the boundary in the Y-direction and multiple distinct planes. Going from left to right in the X-direction, the ascending sides are facets oriented along the SBP, which are comprised of diamond-shaped C units shown in figure 2(b). The C unit is the characteristic unit for Σ11 <110> tilt boundaries in the SUM [21]. The descending side is comprised of a pair of E units (also sometimes referred to as kite-shape structures), which are characterized by a column of free volume in their center [20]. Figure 2(c) shows the E unit pairs outlined in red as well as their location with respect to the C units on the SBP facets. Though E units are at times categorized into two variants [36], one standard (E) and one elongated (E'), the high temperatures and dynamic boundary behavior in this study make distinguishing the two from each other challenging and we will refer to both types simply as E units. We also note that OVITO common neighbor analysis does not always show all six atomic columns as 'other,' such as in the case of the lower E units in figure 2(c). During annealing, the atomic columns that comprise E units dissociate at times, buckling so that there is less free volume at the E unit's center. A schematic example of this process is shown figure 2(d). This process, which we call atomic column dissociation, is a fundamental component of boundary migration in Al.

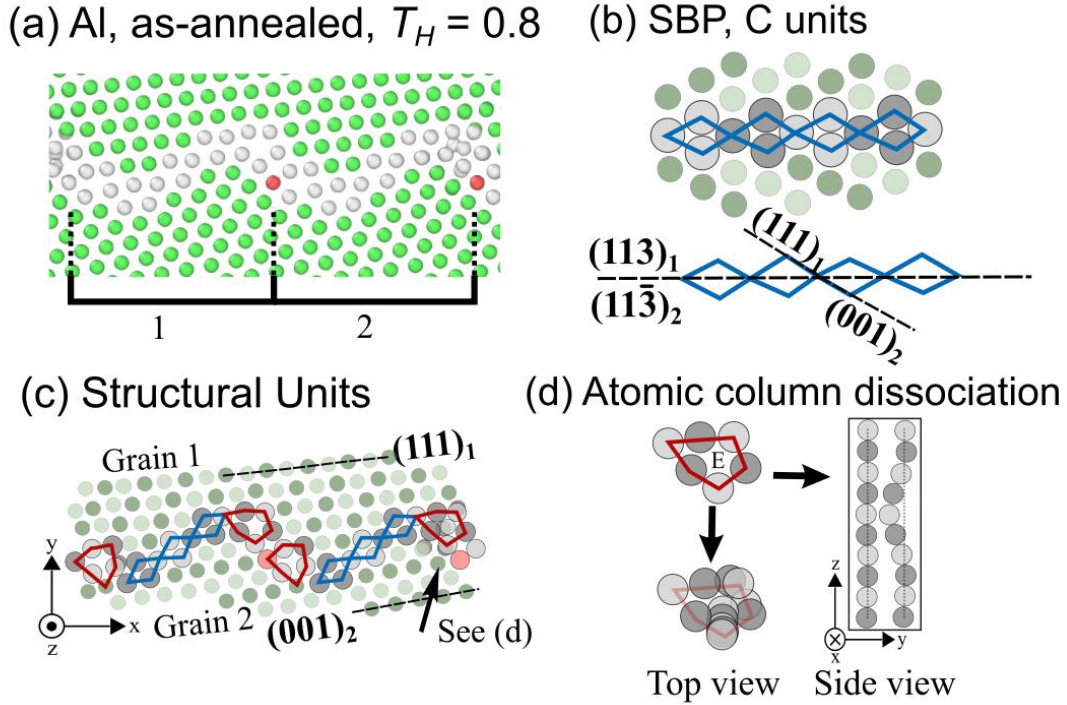

Figure 2. (a) Two facets are shown in more detail from the Σ11 boundary in Al. (b) C units that characterize the Σ11 symmetric boundary. (c) Analysis of the facets, with structural units outlined and important planes indicated. (d) Atomic column dissociation, where one or more columns in an E unit buckles into the free volume at the E unit's center.

Figure 3(a) shows two facets of the as-annealed Σ11 boundary in Cu. These boundaries facet along a different plane orientation, namely an incommensurate boundary plane (IBP) with an orientation of $(111)_1/(001)_2$. The term incommensurate indicates that the ratio of the plane spacings between two component planes is irrational, in this case, $\sqrt{(3 / 1)}$. Since boundaries in CSL systems must have rational plane spacing ratios (such as the SBP, which has a plane spacing ratio of $\sqrt{(11 / 11)} = 1$), this means that an IBP interface by itself could not form a Σ11 grain boundary. In fact, the misorientation needed for an IBP is different than the Σ11 misorientation, requiring the other facet segment to have the correct misorientation. In this case, the SBP facet correctly reflects the misorientation associated with a Σ11 interface. The fact that the IBP appears as a preferred facet plane in this asymmetric Σ11 boundary in Cu is a result of it being locally energetically favorable, even though it would not globally satisfy the misorientation between the grains. Details of the IBP facet structure will be discussed shortly in this section, while information on their energetics, formation, and faceting patterns can be found in other studies, such as those by Brown and Mishin [3] or Wu et al. [6]. At the bracket markings in figure 3(a), it can be seen that IBP facets are separated by one or two $(111)_1$ steps on top and a small defect that impinges into Grain 2 by a few Å. The impinging pattern and the presence of HCP-coordinated atoms (colored red) are evidence of Shockley partial emission from the boundary, which create an array of non-planar defects. It is a relaxation mechanism common to many <110> tilt boundaries in low stacking fault energy materials, including many asymmetric Σ11 boundaries [3, 20, 21]. The presence of Shockley partials was confirmed using the dislocation analysis (DXA) algorithm in OVITO [37], shown in figure 3(b).

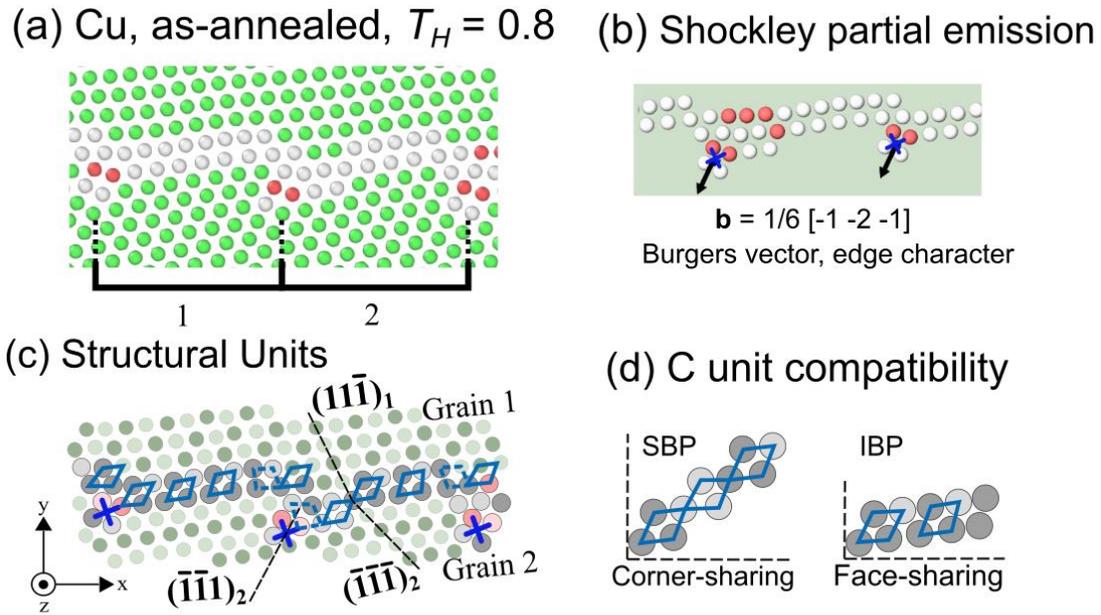

Figure 3. (a) Two facets are shown in more detail for the Σ11 boundary in Cu. (b) Shockley partial dislocations are emitted from facet nodes, with their location indicated by a blue X. (c) Analysis of the facets from (a), with structural units outlined and important planes indicated. (d) C unit compatibility between the symmetric boundary plane facet and the incommensurate boundary plane facet.

Structurally, the IBP has been previously interpreted as a quasi-periodically repeating series of E units [3], and the sites of Shockley origin as a special variant of them, the E'' unit [38]. There is another possible interpretation of its structural units that highlights a special crystallographic relationship between the IBP and the SBP facets which is relevant to migration. The IBP's defining $(111)_1/(001)_2$ planes are the also constituent planes of individual C units from the SBP. This means that IBP facets and SBP facets are crystallographically compatible with each other. Where C units in the SBP facets of Al from figure 2(b) and (c) are corner-sharing, C units in the IBP could be described as face-sharing. The detailed schematic of structural units shown in figure 3(c) shows how these units line up in an IBP facet, and the crystallographic compatibility between corner-sharing and face-sharing C units is highlighted in figure 3(d). This structural relationship can be observed in the different facet variants that populate the as-annealed boundary in Cu. Starting from the facet node at the emitted Shockley, Facet 1 in figures 3(a) and (c) contains 5 face-sharing C units. One of these units has an angular distortion indicated by using dashed instead of solid lines. In its neighbor Facet 2, the Shockley partial has migrated down one $(001)_2$ plane, creating a pair of corner-sharing C units. Facet 1 represents the simplest variant of IBP facet with only face-sharing C units, while Facet 2 has a combination of face- and corner-sharing C units.

To conclude our introduction of Σ11 boundary structure, we address the topic of facet junctions and facet junction defects. The structural complexity of the Σ11 boundaries, the use of two different materials, and the study of dynamic boundary structure (next section) makes the definition of a facet junction somewhat complicated. E unit pairs could be interpreted in several ways, for example as defect-heavy facets by themselves (since they have a relatively clear plane orientation) or as two junction defects, one upper and one lower, that link SBP facets. In the case of the Σ11 boundary in Cu, variations in facet types via C unit unfolding also make the strict identification of the facet junctions complicated. Therefore, instead of defining facet junctions and facet junction defects explicitly, we will instead refer to facet nodes. The dashed lines above the brackets in figures 2(a) and 3(a) show the sites of each facet node. The facet node in Cu is defined as the site between the two $(-1\,-1\,1)_2$ planes surrounding the emitted Shockley partial. The facet node in Al is defined as the site between the two E units. The defects that appear at facet nodes (E units and Shockley partials) will be referred to as facet node defects. The term 'facet' by itself will exclusively apply to IBP and SBP facets during the coming discussion.

### 3.2 Overview of Directionally-Anisotropic Mobility

Figure 4 shows mean grain boundary displacement as a function of time for the six unique simulations at $T_H = 0.8$, for both potentials (rows) and migration types (columns) over a time of 70 ps. Since the ADF magnitude is kept constant, the slope is directly related to mobility, with a steep slope signaling a high mobility. The slopes of both Al boundaries in figure 4(a) and (b) are very similar to each other in magnitude. The slope for Type 1 Cu in figure 4(c) roughly resembles what was observed in Al, but the Type 2 curve in figure 4(d) is completely flat, signaling that the Cu boundary is initially immobile in one direction (when the ADF is applied to Grain 2) within the same time span. Given enough time, the boundaries do eventually move, but with significantly lower velocities (slopes) than those of Type 1. Figure 4(e) shows the two different modes of behavior by examining the displacement of a longer time. On the left side is the immobile phase, which is at least 100 ps long for $T_H = 0.8$. On the right side is the mobile phase, which shows the trajectories used to calculate mobilities (shifted to 0 ps in 'mobile time' to facilitate comparison). The clear difference between the Type 1 slopes for Cu in figure 4(c) and the Type 2 slopes in the mobile phase of figure 4(e) show that the $\Sigma 11$ Cu boundary mobility cannot be uniquely defined by one mobility value, instead requiring two separate mobility values $M_1$ and $M_2$ to accurately capture its behavior.

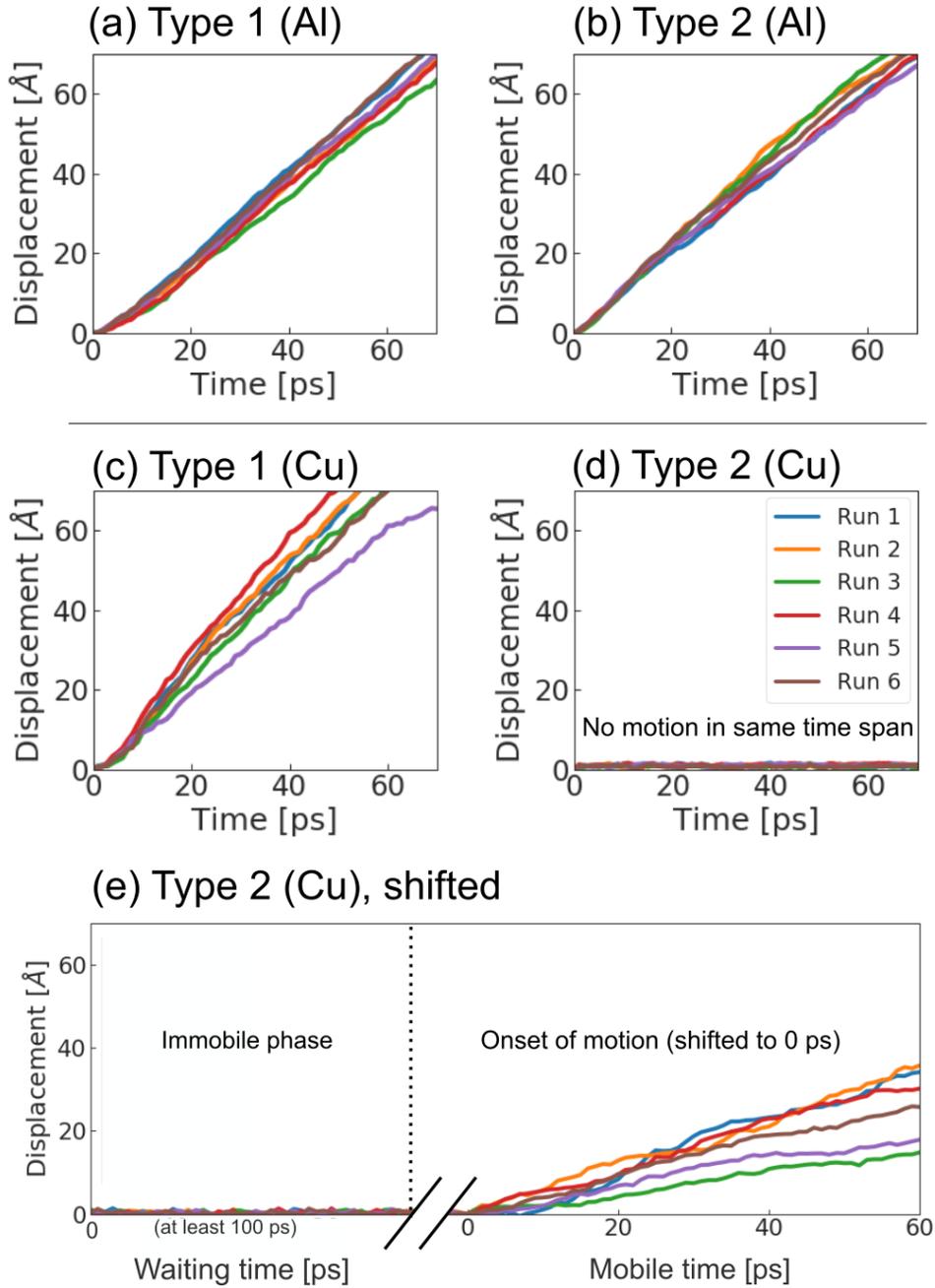

Figure 4. Trajectories of the Σ11 boundary in Al for (a) Type 1 and (b) Type 2 motion at $T_H = 0.8$, providing an example of directionally-isotropic mobility. (c) The trajectory of the Σ11 boundary in Cu undergoing Type 1 motion at $T_H = 0.8$. (d) The Type 2 trajectory for the Σ11 boundary in Cu over the same time shows a lack of migration. (e) Trajectories from longer time simulations of Type 2 motion in the Σ11 boundary in Cu show that the interface eventually moves but with a much lower slope/mobility. This boundary has an immobile phase (left) and a mobile phase (right). The axis is broken to indicate that the waiting times varied for different runs, and trajectories are shifted to the onset of steady-state velocity.

Figure 5 shows the temperature-related mobility trends for both boundaries, first with mobility plotted as a function of temperature in figures 5(a) and (b) and then with the same data replotted in Arrhenius coordinates ($\log(M)$ as a function of inverse temperature) in figures 5(c) and (d). Error bars in all figures show the standard deviation around the mean value. For Al in figure 5(a) and (b) (blue squares), mobility increases with increasing temperature in an essentially identical manner for both Type 1 and Type 2 motion. The Cu boundary (red circles) by contrast has varying temperature trends with motion type. In figure 5(a), the Type 1 first remains relatively constant and then decreases with increasing temperature. In contrast, Type 2 mobility increases with increasing temperature in Cu.

   Though our study does not include a large enough temperature range to thoroughly analyze temperature-related mobility behavior in these faceted boundaries, there are a few trends from $T_H = 0.8$ to 0.9 worth exploring in more depth. The Arrhenius plots in figure 5(c) and (d) allow for a clearer visualization of these trends. The different inverse temperatures for each material are shown on different axes, with Al on top in blue, and Cu on bottom in red. The Al curves in figure 5(c) and (d) are both linearly decreasing Arrhenius curves that that appear to be consistent with thermally activated grain boundary motion [10, 11]. In contrast, the Cu curves have more complex temperature-mobility trends. To quantify these behaviors, we extract activation energy barriers (Q) from the mean values of these Σ11 curves, as well as those from a similar Arrhenius analysis of the Σ5 boundaries (not shown), with the results displayed in Table 2. According to the thermal classification criteria set up by Homer et al. [11], boundaries classified as thermally activated have values ranging from 0.01 to 1 eV. Since the majority of energies fall within this range, most of the boundaries studied here can be classified as thermally activated as well.

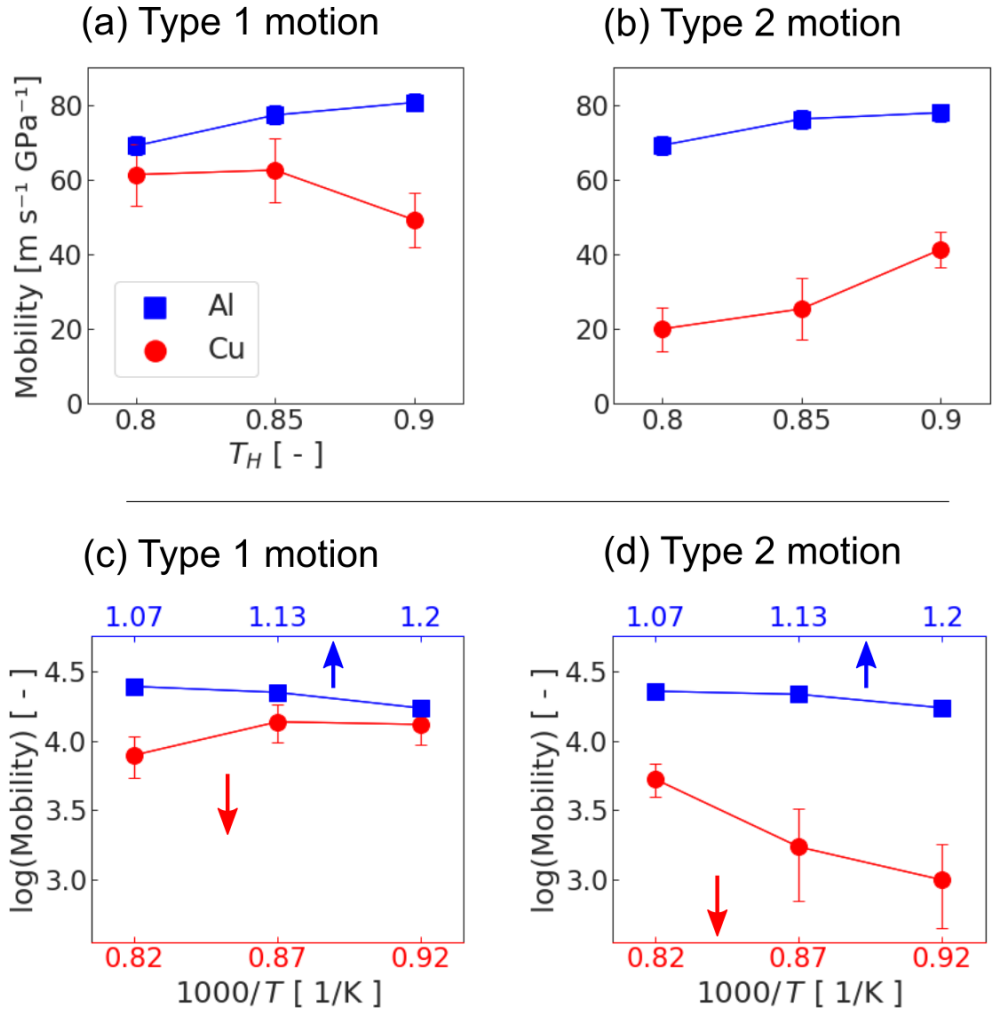

Figure 5. (a, b) Type 1 and Type 2 mobilities at different temperatures for Σ11 boundaries in Al (blue) and Cu (red). (c, d) Arrhenius plots showing the mobility for the same boundaries as a function of 1,000 / T for each material. The bottom (red) and top (blue) axes show the values for Cu and Al, respectively.

| Boundary type | Potential | Migration Type | Q [eV] |
|---|---|---|---|
| Σ11 | **Cu** | **1 (faster)** | **-0.19** |
| | | 2 (slower) | 0.65 |
| | Al | 1 | 0.10 |
| | | 2 | 0.08 |
| Σ5 | Cu | 1 | 0.10 |
| | | 2 | 0.09 |
| | Al | 1 | 0.17 |
| | | 2 | 0.14 |

Table 2. Activation energy barriers (Q) derived from Arrhenius analysis of mobility vs. temperature data for the Σ11 and Σ5 boundaries for each potential. Values of between 0.01 and 1 eV indicate that the boundary is thermally activated [11]. The negative activation energy of the anisotropic Type 1-driven Σ11 boundary in Cu (bold) is consistent with thermal dampening.

The major exception is the anisotropic Σ11 Cu boundary (first two rows of Table 2). Like the other boundaries mentioned above, the Type 2-driven Σ11 boundary in Cu is also thermally activated. However, both of its energy barriers are a great deal higher than all other activation energies observed here. While the Σ11 boundary in Al and the Σ5 boundaries all have activation energies lower than 0.2 eV, the Type 2 boundary has a significantly higher value of 0.65 eV. By contrast, the energy barrier for Type 1 (faster) migration has a negative value of -0.19, characteristic of thermally-damped motion. These contrasting behaviors in the Cu boundary indicate that it is not only directionally-anisotropic with respect to mobility, but also with respect to thermal motion behavior.

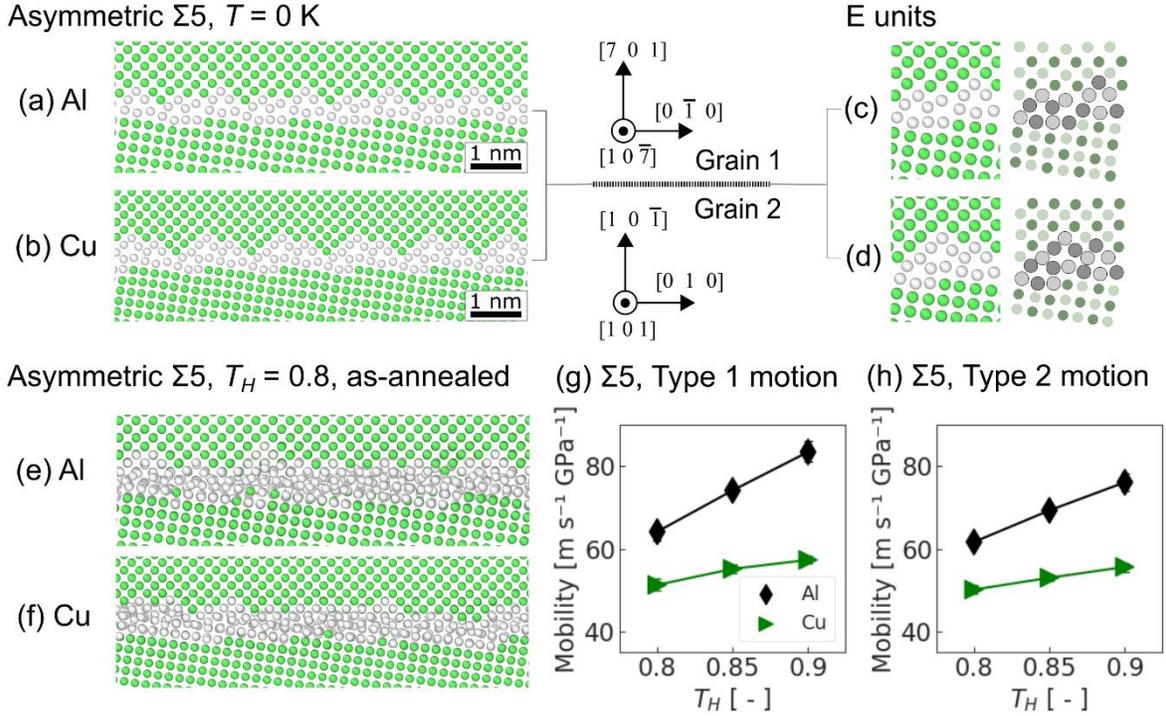

Figure 6. (a, b) The minimized (T = 0 K) asymmetric $\Sigma 5$ boundaries in Al and Cu that are used as baselines for comparison. (c, d) Zoomed views show that both boundaries contain a series of E units. (e, f) The as-annealed asymmetric $\Sigma 5$ boundaries at $T_H = 0.8$, where the structure becomes more disordered at elevated temperature. (g, h) Mobilities of asymmetric $\Sigma 5$ boundaries in Al (black) and Cu (green) as a function of homologous temperature. All boundaries here demonstrate thermally-activated mobility trends.

Though asymmetric tilt boundaries have been a part of many mobility studies (see, e.g., [10-14]), there have been few that mention mobility in different directions specifically [10, 14]. It is important therefore to establish a baseline of what is expected for typical asymmetric boundaries, by comparing the mobility trends of a relatively unremarkable asymmetric boundary to a faceted one. To accomplish this, we chose a non-faceted, asymmetric $\Sigma 5$ <001> tilt boundary, shown at T = 0 K in figure 6(a) and (b) for Al and Cu, respectively. The minimized structures are made of E units/kite-shape structures similar to those of $\Sigma 11$ boundary in Al, as shown schematically in figure 6(c) and (d). $\Sigma 5$ tilt boundaries have a predictable structure while also having energies similar to those of general high-angle interfaces [5, 36], which can be found in Table 1. Figures 6(e) and (f) show the as-annealed structures, where the addition of temperature leads to a qualitative loss of structural definition that the Σ11 boundaries do not undergo at the same homologous temperature, which can be partially explained by their relatively high energies compared to Σ11 boundaries (Table 1) and the interconnectivity of free volume between E units [20]. Simulations for the asymmetric Σ5 boundaries were run using identical parameters to those used for the Σ11 boundaries. In terms of mobility, shown in figure 6(g) for Type 1 and figure 6(h) for Type 2, neither boundary exhibits the large differences that were observed for the Σ11 boundary in Cu. Analysis of temperature-mobility trends using Arrhenius plots (not shown) suggest that all of these boundaries move by a thermally activated mechanism.

To quantify and thus more precisely compare the relationship between Type 1 and Type 2 mobilities, we define the mobility anisotropy ratio, $A$, as:

$$A = \frac{M_1}{M_2}. \hspace{3cm} (2)$$

A mobility anisotropy ratio of 1 indicates that the mobilities of Type 1 and Type 2-driven boundaries are identical. Values higher or lower than 1 than that indicate the factor of increase or decrease for Type 1 mobility relative to Type 2. The mobility anisotropy ratios for the various homologous temperatures and materials are plotted in figure 7. The average anisotropy ratios in the Σ11 Cu boundary at $T_H = 0.8$ and 0.85 are far higher than the other boundaries, with average magnitudes of 3.3 and 2.8. Though this anisotropy is reduced significantly at $T_H = 0.9$, it is still somewhat elevated with a value of 1.2. A zoomed view of anisotropy values below 1.4 is shown in figure 7(b). All of the anisotropy values of the Σ11 boundary in Al and both Σ5 boundaries are below ~1.1. Overall, the Σ11 boundary in Al has the lowest anisotropy values, all lower than 1.05. The data shown in figure 7(b) suggests that a typical or unremarkable asymmetric boundary would have mobilities that vary by 10% or less in the two opposite directions. The large deviation of the Σ11 boundary in Cu is therefore notable. Since the anisotropy values of the Σ11 boundary in Al lie close to 1, and beneath those of the Σ5 Al boundary, its anisotropy values fall within the range expected for a general asymmetric boundary. These findings align with the structural trends already observed above in the as-annealed boundaries shown in figures 1-3. The E unit pairs present in the Σ11 boundary in Al are similar to those seen in the Σ5 boundaries for both potentials. The Σ11 boundary in Cu by contrast has the unique feature of emitted Shockley partials, and a uniquely high mobility anisotropy.

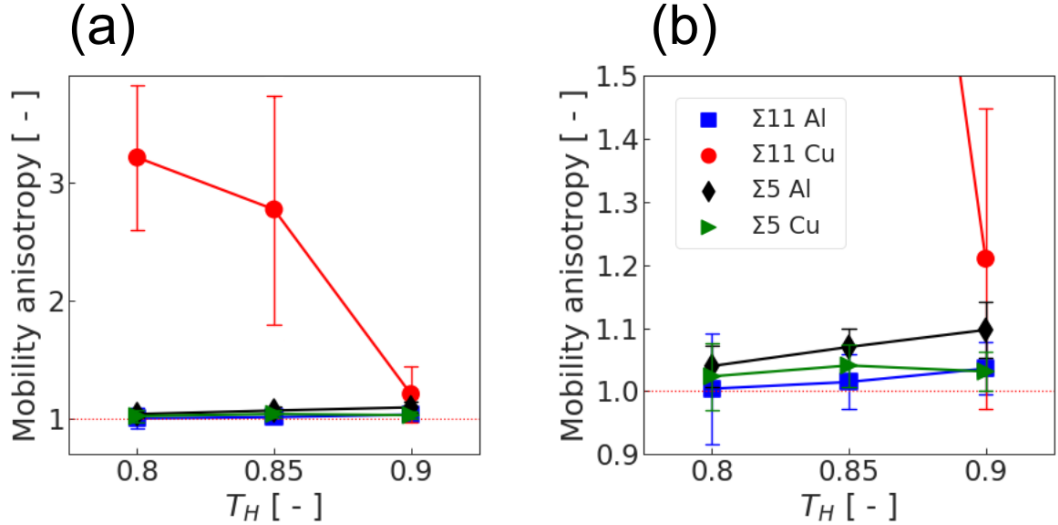

Figure 7. (a) The mobility anisotropy, A, as a function of temperature for both faceted Σ11 boundaries and asymmetric Σ5 boundaries in Al and Cu. (b) A zoomed view of the values with lower anisotropy, around $A = 1$.

### 3.3 Common Migration Mechanisms

In this section, migrating Σ11 boundaries are observed in order to explore possible relationships between boundary structure and mobility anisotropy, with an initial focus on shared mobility mechanisms. Figure 8 presents a snapshot of Σ11 boundaries after the application of the ADF for 65 ps at $T_H = 0.8$. The Type 1-driven and Type 2-driven boundaries in Al in figure 8(a) and (b), respectively, are outlined in blue and the Type 1-driven boundary in Cu in figure 8(c) are outlined in red. The large gray arrows on the right side of each image indicate the direction of the applied driving force. In each image, one representative SBP facet and IBP facet have been labeled for reference. As with the as-annealed boundaries of figures 2(a) and 3(a), facets have been marked with brackets beneath each image and numbered from left to right. Each image is also labeled with the calculated mobility value for the boundary that is shown.

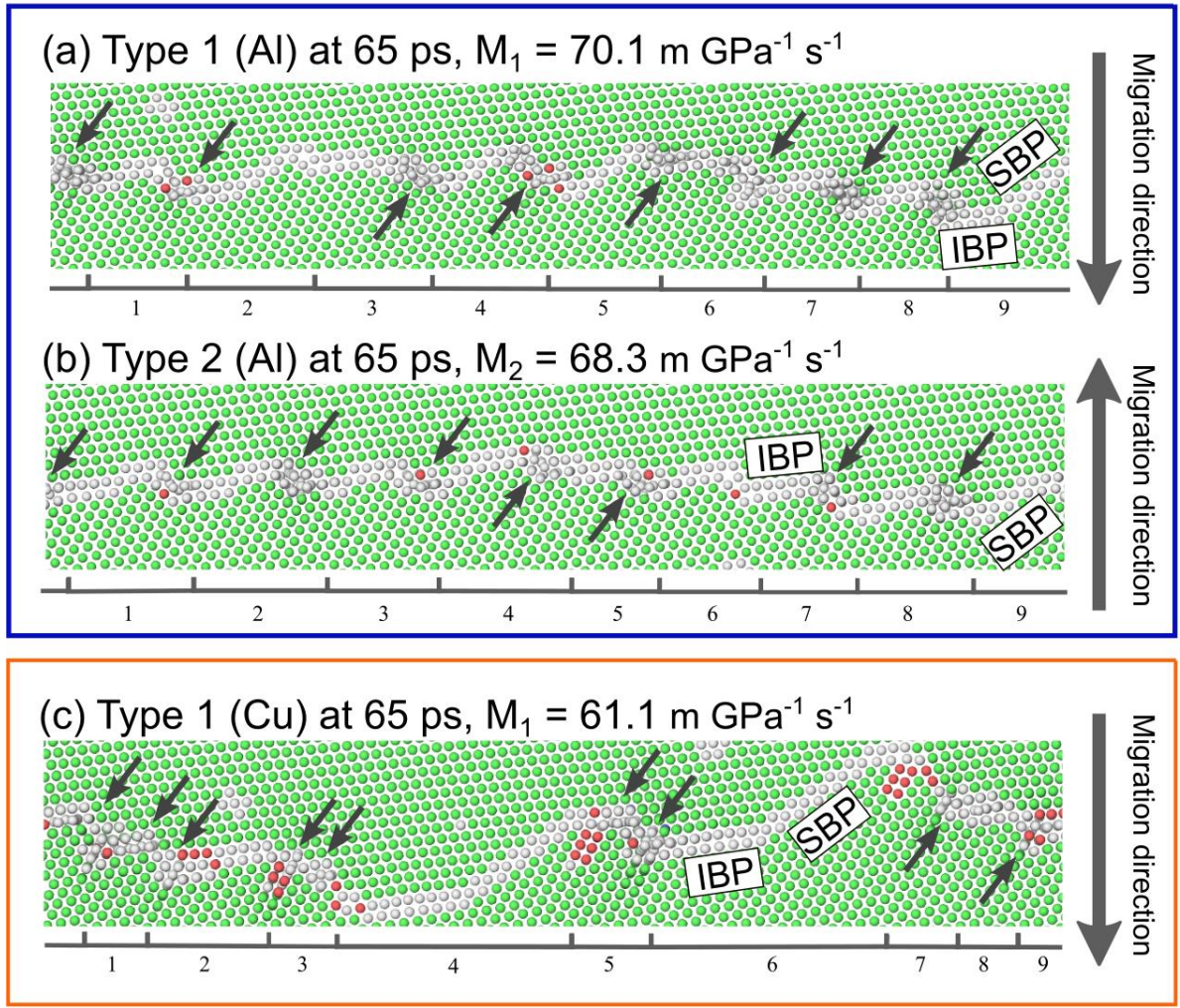

Figure 8. Snapshots of boundary motion at $T_H = 0.8$ for (a) a $\Sigma 11$ boundary in Al undergoing Type 1 motion, (b) a $\Sigma 11$ boundary in Al undergoing Type 2 motion, and (c) a $\Sigma 11$ boundary in Cu undergoing Type 1 motion. These boundaries all move relatively smoothly and have 'normal' migration. The brackets beneath each snapshot indicate the location of facet nodes, and the numbers the respective facet period. The black arrows indicate facet nodes where atomic column dissociation has occurred.

Figure 8 provides an overview of the evolution of structure in migrating Σ11 boundaries. In general, it is the defects at facet nodes, indicated by brackets beneath each boundary snapshot similar to those shown in figures 2 and 3, that provide most of the boundary displacement. Even after significant migration, many features of the as-annealed boundary structures are still recognizable, but structural units originally native to only one element type now appear in both elements with a high frequency. For example, the IBP facets, favored by Cu in the starting structure, are regularly observed in the Al boundaries during both Type 1 and Type 2 motion. Likewise, the Cu boundary has formed two SBP facets that were not present in the starting structure, likely as a consequence of the C unit compatibility between the two facet types mentioned above in figure 3. In addition, the Type 1-driven boundary in Cu has formed several E units similar to those in the Al boundary, whereas it only had emitted Shockley partials in the as-annealed form. The black arrows in figure 8 indicate facet nodes that have undergone atomic column dissociation, shown originally in figure 2(d), and recognizable by the disregistry of atoms along the tilt axis (Z-axis). The dissociation isn't unique to E units but can also be observed at Shockley partial emission sites in the facet nodes of the Cu boundary as well in figure 8(c). By comparing the boundaries of figure 8, we learn that the only phenomenon truly unique to the Type 1-driven boundary in Cu is the emission of Shockley partial dislocations. The atomic column dissociation seen in the as-annealed Σ11 boundary in Al is present at Shockley partial emission sites in the Cu boundary as well (black arrows in all snapshots), recognizable by the disregistry of atomic columns along the tilt axis (Z-direction). The only structures unique to the Type 1-driven boundary in Cu are the emitted Shockley partials.

To better understand mobility anisotropy, two of the most common means of facet node migration are outlined. The first, common to the Σ11 in both Al and Cu and in both directions, is

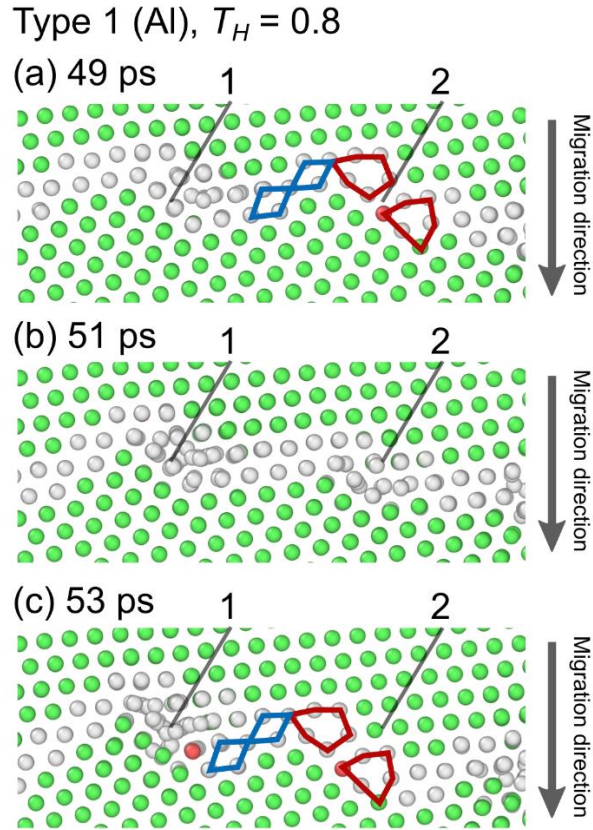

Figure 9. Demonstration of two kinds of disordered shuffling that may occur during facet node migration in the $\Sigma11$ boundary in Al during Type 1 migration at $T_H = 0.8$.

a complex atomic shuffle taking place in and near facet node defects. This shuffling occurs when one or more atomic columns in a facet node defect dissociate as described earlier (figure 2(d)), and when this occurs during migration it could be described as "disordered shuffling." Two common variations of disordered shuffling are shown in figure 9 at two different nodes, labeled Node 1 and Node 2. Tracking Node 1 in figures 9(a)-(c) provides an example of an entire facet node involved in disordered shuffling, which is commonly observed in migrating $\Sigma11$ boundaries in Al. These facet nodes move somewhat slowly but only rarely re-associate into E unit pairs. Their local activity can encourage very slow-moving, non-dissociated E units, such as Node 2 in figure 9(a), to begin atomic column dissociation. Sometimes, as is the case here for Node 2 in figure 9(b) to (c), the dissociation starts on one column but does not propagate further and the node returns to its

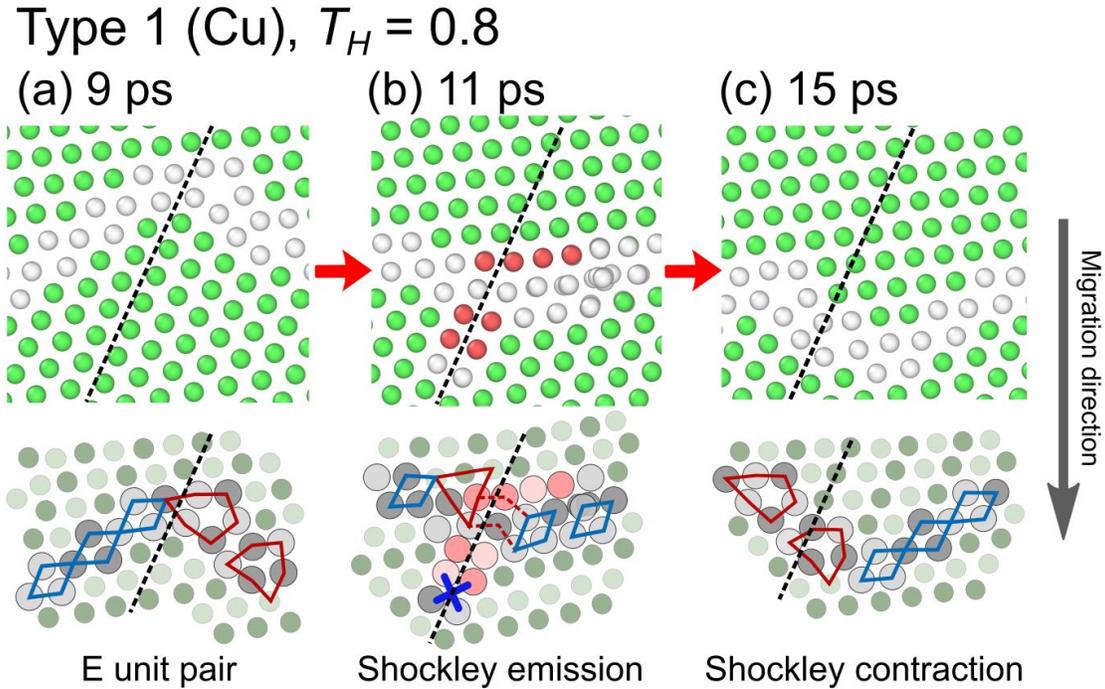

**Type 1 (Cu), $T_H$ = 0.8**

(a) 9 ps  (b) 11 ps  (c) 15 ps

Migration direction

E unit pair  Shockley emission  Shockley contraction original form.  Though the disruption of Node 2 does not result in a longer disordered shuffling process,, the dissociation event in Node 2 does allow it to migrate it one plane height lower into Grain 2 by figure 9(c).

Figure 10.  Demonstration of the Shockley partial emission/contraction process at a facet node (Shockley shuffling) during Type 1 migration of a faceted Σ11 boundary in Cu at $T_H$ = 0.8.

Disordered shuffling is common to all migrating boundaries and is the only means of facet node migration seen in the Σ11 boundaries in Al.  The facet nodes of Σ11 boundaries in Cu migrate using this mechanism as well but also move via cycles of Shockley partial dislocation emission and contraction.  Shockley partial contraction, also called stacking fault constriction, is a process in which an emitted Shockley partial recedes into the interface, forming a new E unit pair in the process.  One full cycle of Shockley emission and contraction at a facet node, which could be called "Shockley shuffling" to distinguish it from disordered shuffling, is shown in figure 10.  The black dotted line indicates the orientation of the (-1 -1 1) plane in Grain 2 along which stacking faults are emitted, and also serves as a reference to mark the initial position of the nodes.  To begin

## Type 2 (Cu), $T_H$ = 0.9

### (a) 30 ps

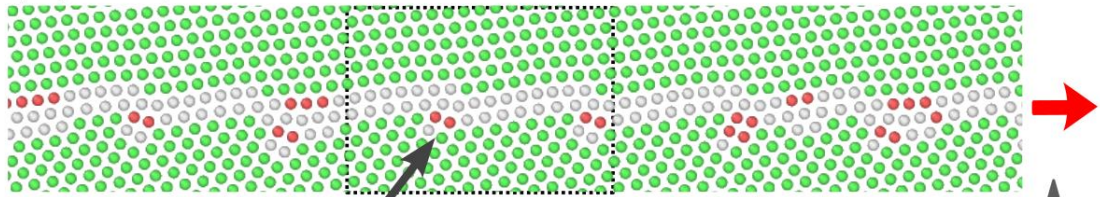

Majority of facet nodes have emitted Shockley partials

### (b) 32 ps

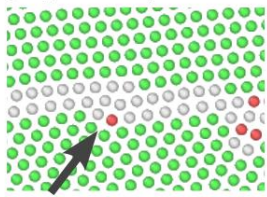

Shockley begins
to contract

### (c) 34 ps

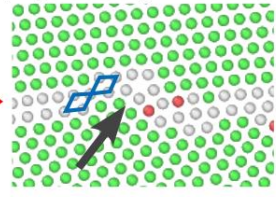

E unit pair forms
C units become
corner-sharing

### (d) 39 ps

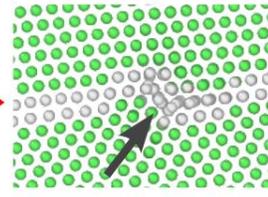

Atomic columns
dissociate

Migration direction

### (e) 45 ps

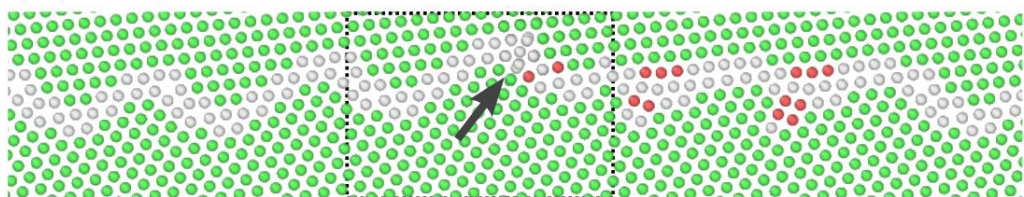

Majority of facet nodes have transformed into E unit pairs a cycle, an E unit must be present, such as that shown in figure 10(a) at 9 ps. At 11 ps, the E unit has emitted a Shockley partial, creating a short stacking fault in Grain 2 (figure 10(b)). Finally, the Shockley partial contracts back into the boundary, re-forming an E unit pair centered around the site of Shockley emission (figure 10(c)). From this point, the E unit may undergo atomic column dissociation and begin moving via disordered shuffling or begin another cycle of Shockley shuffling.  Like disordered shuffling, the stages of Shockley shuffling process are the same for both Type 1 and Type 2 motion.

Figure 11. The transition from the immobile phase to the mobile phase for Type 2 motion in a Σ11 boundary in Cu at $T_H$ = 0.9 is shown. (a) The end of the immobile phase , 2 ps before the the transformation of a Shockley partial into a dissociated atomic cluster (b)-(d). (e) A snapshot of the boundary at the beginning of the mobile phase.

Having understood the general means of facet node defect migration in Σ11 boundaries in Al and Cu, we now move to understand the immobile and mobile phases of Type 2-driven motion in Cu.  Figure 11(a) shows one such boundary at the very end of its immobile phase.  The boundary morphology remains near identical to that of the as-annealed form in figures 1(c), 3(a), and 3(c). In order to begin motion, one or more Shockley partials has to contract into the boundary, as shown in figures 11(b) and (c), resulting in E unit pairs appearing a few step heights above the original site of Shockley partial emission (figure 11(c)).  Once in steady-state motion, as seen in figure 11(e), it takes on the same structures observed in the other mobile boundaries in figure 10, moving via the same Shockley and disordered shuffling  processes at facet nodes.

As can be seen in figure 11(b) and (c), the precipitating event to the beginning of Type 2 migration is the contraction of the Shockley partial dislocation at a single facet node back into the boundary.  This suggests that the length of the immobile phase is correlated with the kinetics of Shockley partial contraction.  Work by Bowers et al. [39] on a boundary similar to the IBP provides experimental and computational data that is useful for understanding the contraction reaction in a boundary with very similar defect morphology.  Using HAADF-STEM, Bowers et al. observed the migration mechanisms of a disconnection traveling along a {001}/{110} incommensurate <110> tilt boundary in Au.  These boundaries are populated by a series of what are described as a "five-fold defects," which in the SUM interpretation would be E (or E') units.  Like the E units at facet nodes in the current work, the E units at certain disconnection sites in the Au boundary could emit Shockley partials to relax, which pin the disconnection until the partial contracted to form an E unit once more.  Inspection of the HAADF-STEM images in the frames before the contraction reaction revealed blurring of atomic columns near the base of E units surrounding the disconnection, which Bowers et al. postulated was due to agitated atoms fluctuating in the local free volume. We note the similarity of this blurring description to atomic column dissociation seen in the boundaries studied in this work. The fact that Bowers et al. hypothesized that these fluctuations were associated with point defect diffusion within the E unit would also be consistent with atomic column dissociation, as we have also observed atomic hopping between grains within the E unit's volume during dissociation.

Bowers et al. also undertook a detailed molecular dynamics study of the emitted Shockley partial's transition to an E unit. They speculated that the observed atomic fluctuations in and around local E units leads to kink nucleation, which in turn initiates Shockley partial contraction. An analysis of the energetic transition path from emission to contraction and back to emission revealed that the contraction occurred at the peak of the energy curve (with an energy barrier of approximately 0.42 eV), meaning that contraction is the rate-limiting process to beginning disconnection migration. Though there are important crystallographic differences between the incommensurate boundary disconnections in Au and the $\Sigma 11$ facet nodes in Cu, we believe that the same basic rate-limiting mechanism operates in our boundary as well. The initial immobile phase preceding migration (shown in figure 4(e)) could then be understood as the time necessary to nucleate the initial Shockley contraction event in a single facet node such as that in figure 11(b). That event then leads to a cascade of structural changes (figure 11(c) through (e)) which could theoretically lower the initial energy barrier for Shockley contraction at other facet nodes.

# Type 1 (Cu), $T_H = 0.8$

**(a) 0 ps**

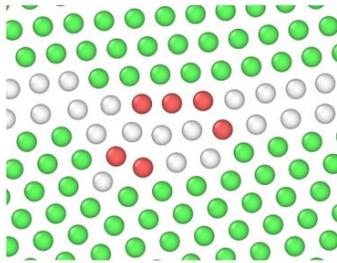

**(b) 1 ps**

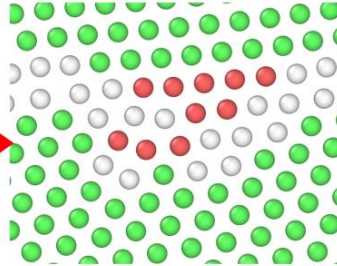

**(c) 4 ps**

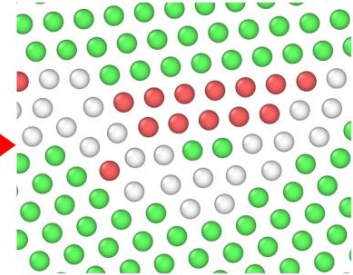

Migration direction

As-annealed

All C units unfold
to corner sharing

Full stacking fault,
E unit pair formed

**(d) 5 ps**

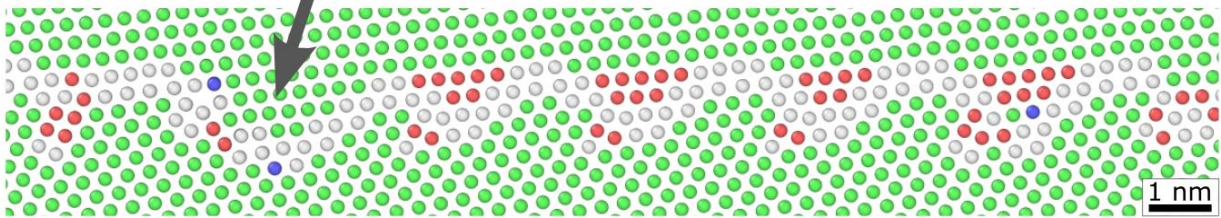

1 nm

**(e) 43 ps**

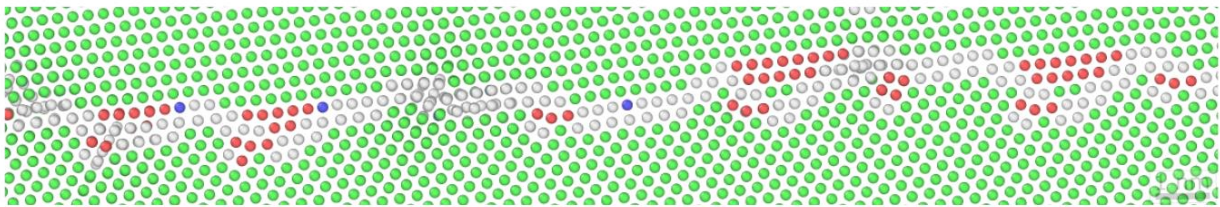

Figure 12. An overview of the process of stacking fault shuffling, which (a)-(e) dominates Type 1 motion in the first 5 ps but (h) also occurs throughout all Type 1 migration simulation runs. Type 1 motion in a $\Sigma 11$ boundary in Cu at $T_H = 0.8$ is shown.

### 3.4 Shuffling modes and directionally-anisotropic mobility

Since Type 1 migration begins with the same array of emitted Shockley partials at facet nodes, it could theoretically also have an immobile phase and the same Shockley partial contraction reaction at a facet node to begin motion. However, Type 1-migrating boundaries have no immobile phase and move on average with a much higher velocity. This suggests that there may be another mechanism operating in this direction of motion. After investigating multiple migrating boundaries in great detail, one can observe that the Type 1 motion in Cu not only migrates via shuffling at the facet nodes, but also takes advantage of the C unit compatibility outlined in figure 3(d) to migrate the facet directly. This process allows the boundary to migrate by replacing an IBP facet with a stacking fault in Grain 1 via C unit unfolding, which in turn is accomplished through small shuffles of each atomic column in the C units. Because the creation of these stacking faults is enabled by the presence of the $(111)_1$ slip plane within the IBP facet, this mechanism could be described as "slip plane shuffling." This mechanism is the primary means of boundary migration in the first 10-20 ps after application of the Type 1 ADF in the $\Sigma11$ boundary in Cu. The initial step of slip plane shuffling is shown in figure 12(a) at the very beginning of a Type 1 simulation. The process continues in figure 12(b) with the unfolding of another C unit, creating a full SBP facet and elongating the growing stacking fault. The third step is the contraction of the Shockley partial into the boundary in figure 12(c), leading to the creation of an E unit pair and SBP facet that are very similar to those observed in the as-annealed boundary in Al. Figure 12(d) shows an example of the high frequency of slip plane shuffling already present at 5 ps. From this point onward, facet migration may proceed in several different ways, through Shockley or disordered shuffling). In most cases, the stacking fault disappears upon further motion (denoted by the black arrow). Slip plane shuffling can also be observed in Type 1 and Type 2 migrating boundaries when moving facet nodes are at approximately the same height with respect to the Y-direction, as shown in the boundary snapshot of figure 12(e) at 43 ps.

Taken together, figures 11 and 12 can provide an explanation as to why slip plane shuffling is not a mechanism available during Type 2 migration in Cu. The directionality of the C unit (which is oriented along the inclination angle $\beta = 35.3°$) restricts the ways in which IBP and SBP facets can connect to each other. Specifically, they both may only act as ascending planes (going from left to right in the figures of this paper) along the $(111)_1$ slip plane of the IBP facet and C unit folding/unfolding must necessarily follow the same rules. As shown in figure 12 above, the unfolding of an IBP facet from left-to-right preserves this ordering. Type 2 migration on the other hand would be attempting to refold the C units in reverse order (going from the state shown in figure 12(c) to the state in figure 12(a)). At that point, since C units in the IBP's center or right side cannot move without breaking up the IBP (and creating forbidden descending facet segments), it is the facet node which must migrate, necessitating a transformation of the node defects. This was observed directly in figures 11(b)-(d), where the boundary is immobilized under the Type 2 ADF until Shockley partial contraction occurs at a facet node.

The slip plane shuffling mechanism can also explain the isotropic mobility of the $\Sigma11$ boundary in Al. This interface has identical boundary asymmetry in terms of C units and IBP facets and must therefore follow the same migration rules. However, its higher stacking fault energy means it does not emit stacking faults. Its mobility is thus limited only by the rate of E unit-based disordered shuffling. This implies that directionally-anisotropic mobility in $\Sigma11$ boundaries in Cu is not a result of Type 2 being slower than Type 1, but rather a result of Type 1 migration being faster. The Type 1-specific slip plane shuffling mechanism allows facet nodes to bypass a Cu-specific rate-limiting facet node migration step that can slow movement, namely Shockley partial contraction.

To confirm that slip plane shuffling is occurring at a rate that is able to influence mobility during steady-state migration, we used the fact that it creates, per facet, more HCP-coordinated atoms than Shockley shuffling does. A typical emitted Shockley partial contains approximately 24 to 56 HCP-coordinated atoms per facet (2 to 4 atomic columns with 14 HCP atoms per column), while facets undergoing slip plane shuffling tend to create between 42 to 154 HCP-coordinated (3 to 11 columns) per facet, depending on the part of the cycle they are in (see figures 12(a)-(c)). Using this reasoning, the statistics of atom types can be investigated to get a sense of how frequently slip plane shuffling is occurring during the simulation. If this kind of shuffling is occurring at an increased rate compared to Shockley shuffling during Type 1 motion, one would expect to observe that the number of HCP atoms is higher than that of Type 2.

Figures 13(a)-(c) show the number of HCP-coordinated atoms for the Σ11 boundaries in Cu for each homologous temperature (columns) and motion type (colors). The first 20 ps after the start of each run is left out to exclude the initial increased rate of slip plane shuffling shown in figure 12(g) and atom counting was only conducted during steady-state mobility regimes. Each plot contains the information of 12 samples (6 bicrystals with 2 boundaries) for each boundary motion type over a span of 50 ps, with the number of HCP atoms measured every 1 ps. Therefore, any number quoted in this figure is an instantaneous measurement of the number of HCP atoms in grain boundaries of the simulation cell. Except for emitted Shockley partials, out-of-boundary HCP atoms were not counted. The top row shows histogram of HCP atom count for each motion type, while the bottom row shows the cumulative distribution functions for this same data. For the two lower homologous temperatures (figures 13(a) and (b)), Type 1 motion (red) results in a wider

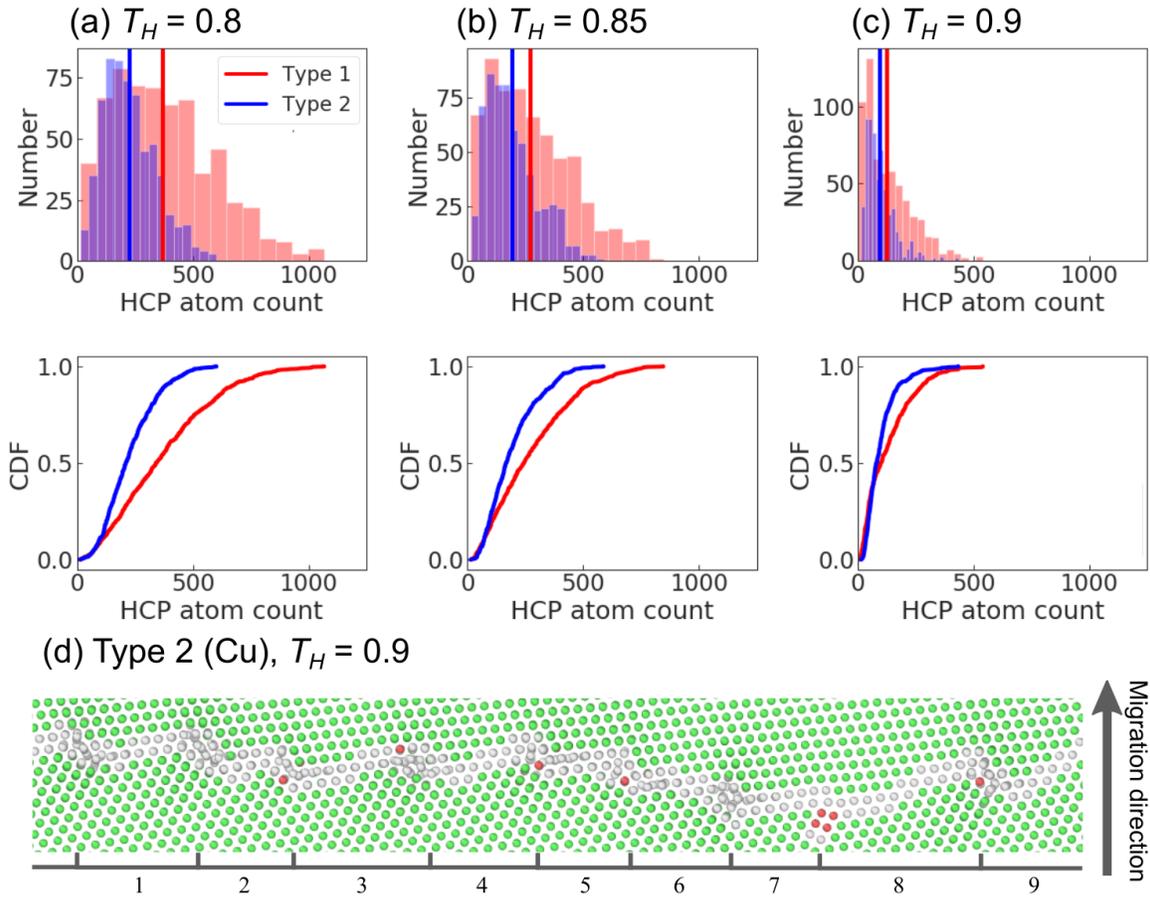

Figure 13. (a-c) Tracking the population of HCP atoms in the simulations at specific times during migration, in both histogram and cumulative distribution function form for Type 1 (red) and Type 2 (blue) motion. The solid colored lines in the histograms show the mean values for each direction. Wider distributions and mean values that are shifted to the right indicate that slip plane shuffling is more active during Type 1 motion. (d) A moving Type 2-driven boundary where Shockley shuffling is suppressed, allowing for rapid migration.

distribution of HCP-coordinated atom counts than that of the Type 2 motion (blue). The mean HCP atom counts for Type 1 migration (red vertical lines) are also higher than those of Type 2 (blue vertical lines). This trend can also be observed in the cumulative distribution functions for Type 1 motion, where the red curves have shifted to the right of the Type 2 curves, indicating that more HCP atoms have appeared on average during Type 1 motion. We can therefore conclude that slip plane shuffling is occurring at a significantly higher rate during Type 1 migration than in Type 2 for the two lowest temperatures studied here. Given the connection between slip plane shuffling and boundary asymmetry, it is likely that this increased occurrence of shuffling is also the source of the higher velocities seen in Type 1 migration versus Type 2. Thus, slip plane shuffling can be identified as the primary mechanism responsible for the directionally-anisotropic mobility in the $\Sigma 11$ boundaries in Cu at $T_H = 0.8$ and 0.85.

## 3.5 Temperature-mobility trends and directionally-anisotropic mobility

Figure 13(c) reveals a possible explanation for the reduction in mobility anisotropy at $T_H$ = 0.9. At this temperature, the distribution of HCP atoms during Type 1 motion undergoes a noticeable shift when compared to the two lower homologous temperatures. Overall, the number of HCP-coordinated atoms shifts towards the lowest range of the histogram and the Type 1 and Type 2 curves in the cumulative distribution function beneath it begin to overlap. The boundary snapshot in figure 13(d) shows that the majority of facet nodes consist of E unit pairs with atomic column dissociation. In other words, the nodes in Cu at the highest homologous temperature resemble those of the $\Sigma 11$ boundary in Al at lower homologous temperatures, which has little to no mobility anisotropy and no Shockley shuffling. The stark decrease in Shockley partial emission at facet nodes due to increased temperature is consistent with what was physically observed in the hybrid HAADF-STEM/molecular dynamics study of disconnection motion by Bowers et al. [39] discussed above in Section 3.3. Recall that an analysis of Shockley contraction in the incommensurate Au boundary supported the idea that this feature is nucleated by point defect diffusion within the free volume of E units. Since increasing temperature also increases point defect diffusion, energy barriers to Shockley contraction are more rapidly overcome. Additional evidence for an increase in Shockley partial contraction (and decrease in Shockley partial emission) can be found when comparing the cumulative HCP atom counts for both Type 1 and Type 2 migration at each temperature, or the area under the curves. The areas of each histogram plot become smaller with increasing temperature, indicating a decrease in the total number of HCP atoms counted during migration. Because the number of $\Sigma11$ facet nodes remains constant, we conclude from this data that Shockley and slip plane shuffling are suppressed. This makes disordered shuffling the dominant migration mechanism in the $\Sigma11$ boundary in Cu (as it is in the $\Sigma11$ boundaries in Al), leading in turn to a corresponding reduction in the mobility anisotropy.

The structural transitions from Shockley and slip plane shuffling modes to disordered shuffling can also explain the directionally-dependent temperature-mobility trends in the $\Sigma11$ boundary in Cu noted above in figure 5 and Table 2. Recall that the activation energy barriers from Table 2 revealed Type 1 motion to be thermally damped (decreasing mobility with increasing temperature) and Type 2 motion to be thermally activated. One possible explanation for both trends could be that the transition to disordered shuffling represents a "facet node roughening" analogous to the well-documented phenomenon of boundary roughening [40-42]. Above the 'roughening temperature' $T_R$, boundaries generally see an increase in their mobility, which is what is observed for the Type 2-driven boundary to occur by $T_H = 0.9$. However, roughening can also lead to decreases in mobility like that observed in a variety of thermally-damped and antithermal boundaries [16, 32, 43, 44]. Studies of the dynamic structures of these interfaces uncover highly ordered atomic shuffling mechanisms that enhance mobility [43]. In those cases, thermal roughening leads to a decrease in mobility when the ordered atomic shuffling becomes disrupted. Similarly, increased temperatures lead to the loss of the ordered, IBP-based slip plane shuffling mechanism in the Type 1 motion direction in Cu, causing it to also lose its mobility advantage over Type 2 (which only has Shockley and disordered shuffling at facet nodes). The intriguing similarities between atom-level shuffling activity on slip planes (such as microrotation around CSL atoms ([43, 44]) and shuffling involving entire slip planes and Shockley dislocations (such as the shuffling in the faceted Σ11 boundaries in Cu, or the antithermal boundaries explored by Humberson et al. [1, 17]) invite future exploration of the role of {111} planes in asymmetric boundary mobility.

## 4. Summary and Conclusions

In this work, molecular dynamics simulations were used to uncover the phenomenon of directionally-anisotropic mobility in a faceted <110> tilt Σ11 boundary in Cu with an inclination angle of β = 35.3°. By comparing its features to boundaries with isotropic mobility, namely a faceted Σ11 boundary in Al and asymmetric Σ5 boundaries in both Cu and Al, the following conclusions can be drawn:

- Asymmetric Σ11 boundaries in Cu can exhibit clear variations in boundary mobility depending on the direction of migration. Motion in one direction was found to be up to three times slower than migration in the other direction. In addition, an immobile phase that was characterized by a long time lag before migration began was observed in many of the slow boundaries. The different motion directions also exhibit different temperature-mobility trends.

- Faceted Σ11 boundary structures, both when stationary and while in motion, can be characterized using only two structural units: (1) C units and (2) E units. The Cu potential with its lower stacking fault energy also emits Shockley partials, which can contract to form the same E unit pairs seen in the facet nodes of the Σ11 boundary in Al.

- SBP and IBP facets are comprised of C units with different relative alignments. This makes boundary transformations between SBP and IBP facets relatively easy, but with a strict orientation set by the inclination angle, β.

- Facet node migration is crucial to boundary migration processes for faceted Σ11 boundaries in both Al and Cu. E unit pairs, which appear in both potentials, can undergo atomic column dissociation, which then moves the facet node (disordered shuffling). Cycles of Shockley partial emission/contraction (Shockley shuffling) may also occur in the Σ11 boundaries in Cu.

- The mechanism of slip plane shuffling is a facet migration mechanism unique to Type 1 motion in the Σ11 boundary in Cu, which arises from the orientation of C units and the compatibility of C units shared between the SBP and IBP facets. This shuffling mechanism provides an explanation for the pronounced directionally-dependent mobility observed at $T_H = 0.8$ and 0.85.

- The magnitude of the mobility anisotropy ratio $A$ is much smaller in the Σ11 boundary in Cu at $T_H = 0.9$. We conclude that this is caused by thermal roughening at facet nodes, which increases the rate of Shockley contraction and appears to also inhibit slip plane shuffling. Without slip plane shuffling, the "fast" boundary becomes slower. The roughening simultaneously increases the mobility of the slower boundary, leading to a significant drop in $A$..

The directionally-anisotropic mobility observed in this faceted Σ11 boundary in Cu underscores the need for atomistic-level study of grain boundary migration. This anisotropy arises directly from the atomic structure of the boundary, motivating a deeper exploration of faceted boundaries in general. The role of {111} grain boundary planes in faceting and mobility could be a particularly fruitful topic of future study. Understanding the impact of unusual migration behavior such as antithermal/athermal mobility trends and anisotropic mobility on microstructural evolution could provide useful insights into phenomena such as abnormal grain growth.

**Acknowledgements**

The authors would like to express their gratitude to Dr. Shawn P. Coleman and the U. S. Army Research Laboratory for providing the ECO force code. This work supported by the National Science Foundation through a CAREER Award No. DMR-1255305.